\documentclass[12pt,preprint]{aastex}

\def\gee{ \, \lower 1mm\hbox{$\,{\buildrel > \over{\scriptstyle\scriptstyle\sim} }\displaystyle \,$}}
\def\lee{ \, \lower 1mm\hbox{$\,{\buildrel < \over{\scriptstyle\scriptstyle\sim} }\displaystyle \,$}}
\def\|{\partial}
\def\o {\over}
\def\Oo {\displaystyle}
\def\varkappa {{\scriptstyle\partial}\! e}

\begin{document}

\title{Bending instability in galaxies: the stellar disk thickness and
the mass of spheroidal component}

%\author{N.V. \surname{Tyurina} \email{tiurina@sai.msu.ru},
% A.V. \surname{Khoperskov} \email{khopersk@sai.msu.ru},
%D.V. \surname{Bizyaev} \email{dmbiz@sai.msu.ru}}

\author{N.V. Tyurina\altaffilmark{1}}
\email{tiurina@sai.msu.ru}
\author{A.V. Khoperskov\altaffilmark{2}}
\email{khopersk@sai.msu.ru}
\and  \author{D. Bizyaev\altaffilmark{1,3}}
\email{dmbiz@sai.msu.ru}

\shorttitle{Bending instability in disk galaxies}
\shortauthors{Tyurina, Khoperskov, Bizyaev}

\altaffiltext{1}{Sternberg Astronomical Institute, Moscow, 119899, Russia}
\altaffiltext{2}{Volgograd State University, Volgograd, 400062, Russia}
\altaffiltext{3}{Physics Department, University of Texas at El Paso, TX, USA}

%-\runningauthor{N.V. Tyurina, A.V. Khoperskov, D.V. Bizyaev}
%-\runningtitle{Bending instability galaxies}
%-\institute{Moscow State University, Russia}

\date{May 06, 2004}

\begin{abstract}

We present results of numerical N-body simulations of a galactic stellar
disk embedded into a spherical dark halo.
The non-linear dynamics of bending instabilities
developing in the disk is studied. The bending modes, axisymmetric and not,
are considered as main factors increasing the disk thickness.
The relation between the disk vertical scale height and the
halo+bulge-to-disk mass ratio is inferred. The method of estimation of the
spherical-to-disk mass ratio for edge-on spiral galaxies based on this
relation is studied and applied to constrain the spherical subsystem mass
and the mass of dark halos ($M_{h}/M_{d}$) in seven edge-on galaxies.
The values of $M_{h}/M_{d}$ are of order 1 for our galaxies.

\end{abstract}


%-\end{opening}

%*********************************************************
\section{Introduction}
%*********************************************************

The luminous matter represents only a part of overall mass
in galaxies. One of arguments for massive dark
matter component comes from numerical simulations of galaxies.
Thus, numerical models evaluated for disk galaxies with light spheroidal
subsystem show that their stellar disks heat themselves
significantly during the evolution achieving the
equilibrium state with too high velocity dispersion
\citep{Ostriker, Carlberg-Sellwood1985, Athanasoula-Sellwood-1986,
Bottema1997, Fuchs-Linden-1998}, \\ \citep{AVH2003}.
To explain the low ratio of velocity dispersion to the gas rotational
velocity $c^{obs}/V_c$ in outer parts which is often observed in many
galaxies, one should consider a spherical subsystem as massive as the
galactic disk in its optical limits.

The vertical scale height of stellar disk depends on the local disk
surface density and the spheroidal subsystem mass $M_s$.
Then, for the case of edge-on galaxies we are enabled to include the
observed disk thickness into the numerical simulations. The disk
thickness is available from observations for many edge-on spiral
galaxies. On the other hand, if the velocity dispersion is close to
the required for the marginal stability of the stellar disk, its
thickness is tied up with the mass ratio of disk to the spherical
component \cite{Zasov1991, AVH20012}.

An important mechanism of the vertical velocity dispersion $c_z$
growth is the bending instability. The bending instability in
galactic disks has been considered with the help of N-body simulations.
Important results were obtained when the tidal interactions
were taken into account \cite{Hernq, Weinberg, Velazquez, Mayer,
Bailin, Reshetnikov2002}.
The conditions for emerging the bending instabilities were
investigated by \cite{Raha, Sellwood, Sellwood-Merritt-1994,
Patsis-2002, Griv-Yuan-Gedalin-2002, Binney}, \\ \citep{Sotnikova}.

In this paper we point our attention to the radial
distribution of $c_z/c_r$ and to the thickness of the stellar
disk needed to provide its stability against different
kinds of bending perturbations. We consider the late
type galaxies without a prominent bulge because the
rotation curve in very central regions is not surely
defined. The presence of bulge would complicate the
analysis of bending instability. The bulge plays a stabilizing role
for bending instability when all other conditions being similar.
If no bulge presents, one can restore the internal part of
rotation curve assuming it is defined by the disk component in
central ($r\lee 2L$) regions \cite{Zasov03}. Here $L$ denotes
the exponential scale length of the stellar disk.

We performed numerical N-body simulations
for seven edge-on galaxies with known photometric scales, both
radial and vertical.
% We include four bulgeless galaxies studied by \cite{dmbiz2002}
% (UGC~6080, UGC~9556, UGC~9422, and NGC~4738). The radial
% distribution of their stellar disks thickness is
% available. The galaxy UGC 7321 is interesting for us
% because of its superthin disk and low surface brightness
% nature \cite{Matthews2000}. For two large and nearby
% galaxies, NGC~891 and NGC~5170, the data on the radial
% component of stellar velocity dispersion are available
% from published data. It enables us to utilize those data
% together with the photometric disk scales.

Our basic interest throughout the paper is to follow up the
evolution of the disk thickness and the behaviour of
$c_z/c_r$ ratio that holds the stellar disk in stability against
the bending perturbations.

%*********************************************************
\section{Dynamical modeling of edge-on galaxies}
%*********************************************************

We evaluated free parameters of the model: $\mu = M_s /
M_d$, the radial scale of the halo $a_h$, and the disk
central surface density $\sigma_0$, looking for the
optimal agreement between the calculated and observed
values of the disk thickness. Here $M_d$ is the galactic
disk mass, $M_s = M_h + M_b$ is the mass of the spherical
component which comprises of halo and bulge in general
case. Note that since almost all our galaxies have no
visible bulge, their spherical component means the
galactic dark halo.

Our dynamical modeling is based on the numerical
integration of the motion equations for $N$
gravitationally interacting particles. This system of
collisionless particles forms a disk embedded into
dark halo and bulge. The steady state distribution of mass
in the bulge $\varrho^{(b)}$ and halo $\varrho^{(h)}$ is
defined as:

\begin{equation}
\label{densityhalo} \varrho^{(h,b)}(\xi) =
{\varrho_{0}^{(h,b)}\o (1+\xi^2/a_{(h,b)}^2)^k} \,,
\end{equation}

\noindent where $\xi=\sqrt{r^2+z^2}$, and $r,z$ are the
radial and vertical coordinates, $k$ equals to 3/2 for the
bulge and to 1 for the halo. The dimensional spatial
scales for the bulge and halo are denoted as $a_b$ and
$a_h$, respectively. We supposed that the bulge is
encompassed by a sphere with radius of $\xi \le
r_b^{\max}$.

The initial vertical equilibrium of the disk is defined by
the Poisson equation

\begin{equation}
\label{Puasson} {\|\o r\| r}\, \left( r{\| \Phi\o \| r}
\right) + {\|^2 \Phi\o \| z^2} = 4\pi G \, \left(\varrho +
\varrho^{h} + \varrho^{b} \right) \,
\end{equation}

\noindent and by balance of forces in the vertical direction in first
approximation

\begin{equation}
\label{z-equilibrim} c_z^2\, {\| \varrho\o \| z} = -
\varrho \, {\| \Phi\o \| z} \,.
\end{equation}

\noindent Here $\Phi$ is the gravitational potential,
$\varrho$ is the disk volume density, and $c_z$ is the
vertical component of velocity dispersion. The  radial
component of the Jeans equation defines the rotation
velocity in the stellar disk \cite{Valluri-1994}:

\begin{equation}
\label{VelocityRotation}
\begin{array}{l}\displaystyle
V^2 = V_c^2 + c_r^2\, \left\{ 1 - {c_\varphi^2\o
c_r^2} + {r\o \varrho c_r^2}{ \| (\varrho c_r^2) \o \| r}
+ {r\o c_r^2}{\| <{uw}> \o \| z} \right\} \,,
\end{array}
\end{equation}

\noindent The last term in the figured brackets in
(\ref{VelocityRotation}) is the chaotic part of the radial
$u$ and vertical $w$ velocity components. We distinguish
between the circular velocity $V_c=\sqrt{ r \left( {\|
\Phi / \| r} \right)_{| z=0}}$ and the stellar rotation
curve $V(r)=r\Omega$ because of the velocity dispersion.

The system of equations
(\ref{Puasson})--(\ref{z-equilibrim}) can be reduced to
the equation for the dimensionless disk density
$f(z;r)=\varrho(z;r)/\varrho(z=0;r)$ \cite{Bahcall}:

\begin{equation}\label{Bachall}
{d^2 f \o dz^2} + 2{d\ln c_z\o dz}\, {df\o dz}-{1\o f}\,
\left( {df\o dz} \right)^2 + {4\pi G\o c_z^2} {\sigma\o 2
z_0}\, f\, \left( f+{2\varrho_h^{eff}\o
\sigma}z_0\right)=0 \,.
\end{equation}

\noindent where the surface density is equal to
$\sigma=2\varrho(z=0)\cdot z_0$, $\Oo
\varrho_h^{eff}=\varrho_h-{1\o 4\pi G r}{\| V_c^2\o \| r}$,
and the vertical scale of the disk is defined as
$z_0=\int^{\infty}_{0}fdz$. The circular velocity is
defined in the galactic plane $z=0$. For the case of
$c_z={\rm const}$ and $\varrho_{h}^{eff}=0$, the disk volume
density profile in the vertical direction is:

\begin{equation}
\label{cosh2} \varrho(z)=\varrho(z=0)\cdot {\rm
sech}^{2}(z/z_0) \,,
\end{equation}

\noindent where we designated the vertical disk scale
as $z_0=c_z^2/\pi G\sigma$. We suppose that the
density distribution in a galaxy follows the
brightness distribution. It corresponds to the
constancy of the mass-to-luminosity ratio. We consider
two possible laws for the vertical density
distribution $\varrho(z)$: $\exp(-z/h_z)$ and ${\rm
sech}^{2}(z/z_0)$. The values of $h_z$ and $z_0$ are
computed as functions of time and location in disk.
The radial density profile in the disk is assumed to
be exponential $\varrho(r) = \exp(-r/L)$. The mass of
the dark halo $M_h$ is calculated inside the maximum
disk radius $R_{\max}$. As a rule,  $R_{\max} \approx
4\,L$ \cite{Kruit1981, Kruit1982, Pohlen, Holley}.

We assume this kind of axysimmetric models in the
equilibrium state as initial templates for our simulations.
The initial radial distributions of the radial ($c_r$)
and azimuthal ($c_\varphi=c_r\varkappa/2\Omega$) velocity
dispersions keeps the disk in gravitationally stable state.

Considering edge-on galaxies, we encounter a well
known problem that their observed structural parameters
are averaged along the line of sight. Once we compare the
model parameters with observations, this effect has to be
taken into account when we reproduce the model rotation
curves as well as the radial and vertical photometric
profiles.
Deriving the surface density, we integrated the volume density of the model
disk along the line of sight. The same integration was performed for the
rotation curves as well, see also \cite{Zasov03}.

%*********************************************************
\section{ The bending instabilities in stellar disks}
%*********************************************************

The disk scale height depends on the vertical velocity
dispersion $c_z(r)$, and the latter is dependent of the radial
velocity dispersion $c_r(r)$. A discussion about the
relation between $c_r$ and $c_z$ was opened firstly by
Toomre \cite{Toomre1966}.
Considering a simplified model of an infinitely thin
uniform self-gravitating layer, \cite{Poliachenko77} found
that an essential condition for the system stability
against the small-scale bending perturbations is $c_z/c_r
\ge 0.37$.

The dynamics of the bending instabilities taking a nonuniform
volume density distribution in $z$-direction into account was
also considered by \cite{Araki-1986} where a lower value of the
critical ratio $c_z/c_r \gee 0.3$ was inferred, see also
discussion in \cite{Merritt}.
The linear analysis of stability against global bending perturbations
was discussed by \\ \cite{Poliachenko79} and generalized by
\cite{Vandervoort}.

%*********************************************************
\subsection{ The bending instability}
%*********************************************************

Let's consider the dynamics of global bending
perturbations. The stellar disks at the initial moment ($t=0$)
were: 1) axisymmetric, 2) in an equilibrium state along
the radial and vertical directions, 3) gravitationally
stable in the disk's plane (it was made by assuming a high
value for the radial velocity dispersion $c_r \gee
(1.5\div 4)\cdot c_T$, the ratio $Q_T=c_r/c_T$ is a
function of $r$ \cite{AVH2003}. The disks could be
either stable or unstable against the bending
perturbations in dependence of $c_z(r)$ distribution. The
scale height $z_0$ and velocity dispersion depend of each
other \cite{Bahcall}, hence lesser value of $c_z$
corresponds to a thinner disk.

The evolution of systems with a small initial value of
$c_z/c_r$ (i.e. the system that is dynamically cold in the
vertical direction) reveals a gradual growth of the global
bending instability which heats the disk up in the
vertical direction, and rises its thickness.

\begin{figure}[!t]
{\includegraphics[width=0.33\hsize]{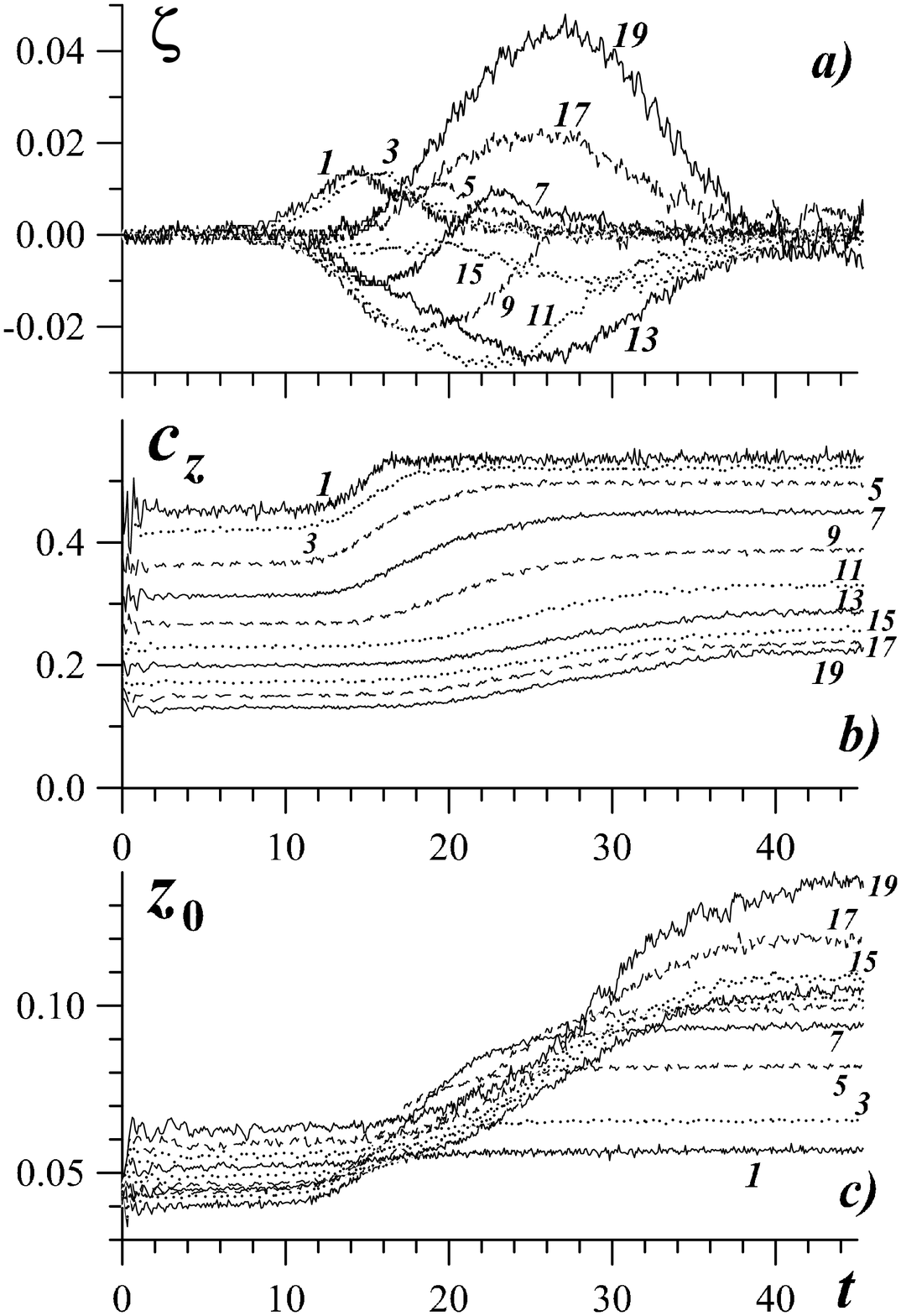}}
 \vskip -0.49\hsize \hskip 0.34\hsize
{\includegraphics[width=0.33\hsize]{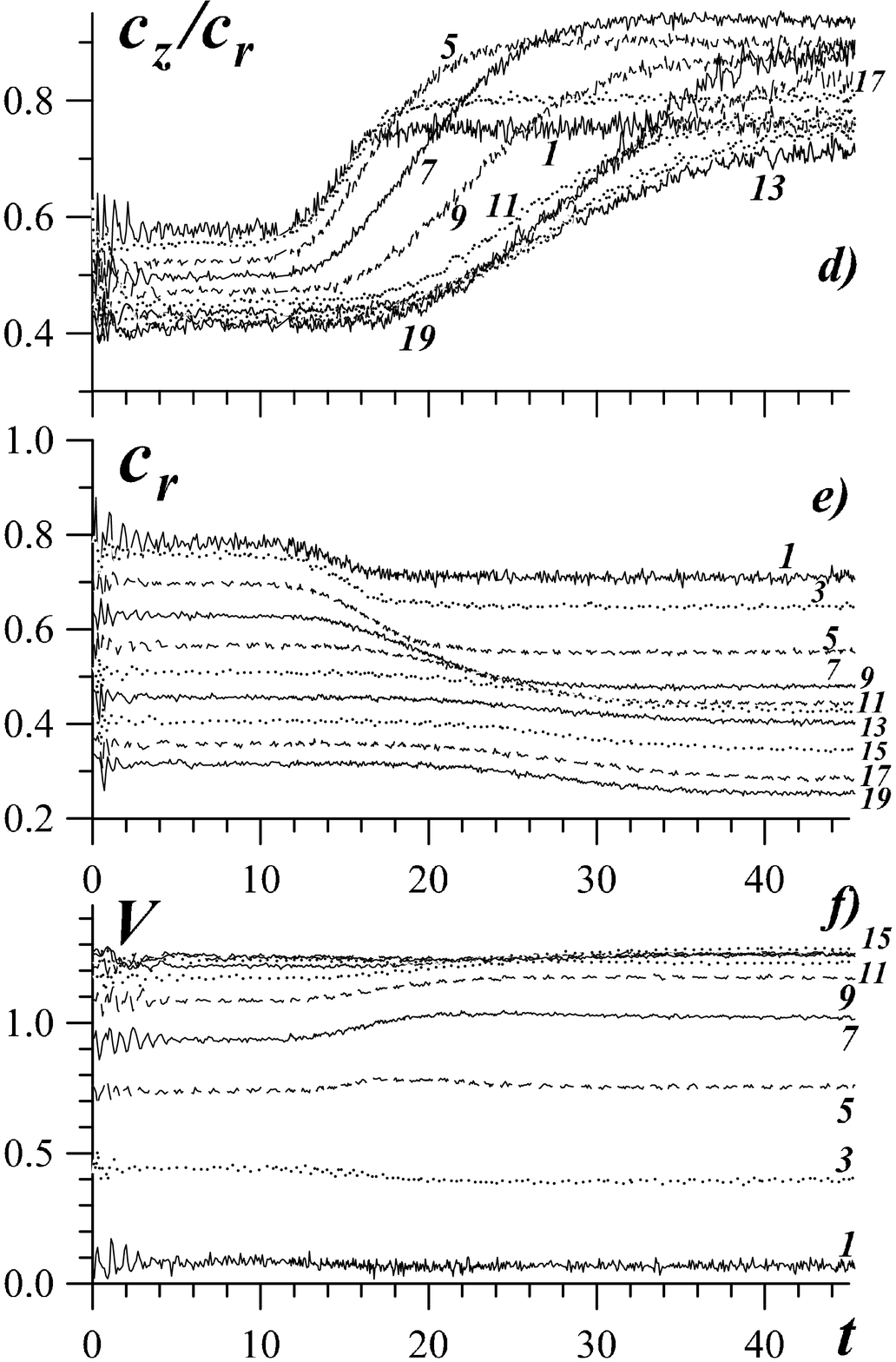}}
 \vskip -0.49\hsize \hskip 0.67\hsize
{\includegraphics[width=0.33\hsize]{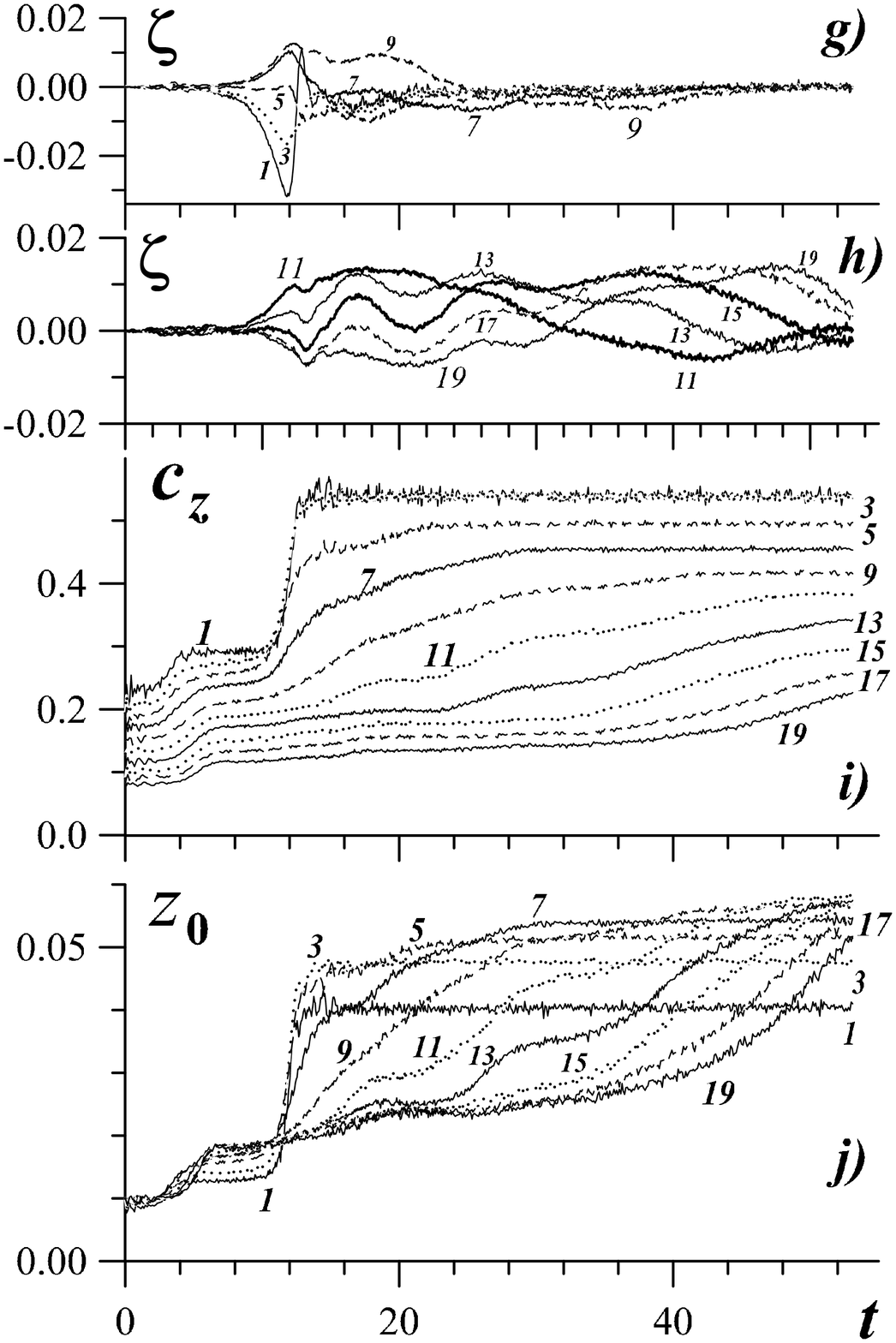}}\vskip
-0.01\hsize
  \caption {
Evolution of the stellar disk parameters when the axisymmetric
bending mode being developed in the model with $\mu=1$,
$a=L$: a) the vertical coordinate of the disk's local barycenter
$\zeta$, b) vertical component of the velocity dispersion
$c_z$, c) disk's scale height $z_0$, d) the vertical to
radial velocity dispersion ratio $c_z / c_r$, e) $c_r$, f)
rotational velocity of particles in the disk $V$.
 g) --- i) The same for the model with $\mu = 4$.
The vertical coordinate of barycenter $\zeta$ for the internal and
external regions is shown in separated plots g) and h)
respectively. Different curves are drawn for a set of
distances to the center $r_j = 4L\cdot(0.05j-0.025)$, where $j$ is
shown in the plots.
All the parameters are averaged by the azimuthal coordinate.
}
\label{Bend-B-Fig1}
\end{figure}

{\it The bending mode $m=0$}.
The developing of unstable axisymmetric bending mode
($m=0$) leads to the significant disk heating in the
vertical direction \cite{Sellwood}. Let's consider the
stellar disk where the axisymmetric mode is developing.
The evolution of the vertical coordinate of the disk's local barycenter
$\zeta(r,t)$, the vertical velocity dispersion
$c_z(r,t)$, the disk scale height $z_0(r,t)$, the
dispersion ratio $\alpha_z\equiv c_z(r,t)/c_r(r,t)$, the
radial velocity dispersion $c_r(r,t)$, and the rotational
velocity $V(r,t)$ is shown in Fig.\ref{Bend-B-Fig1}. We
fix the units by the assumption $M_d = 1$, $G=1$, $4L=1$.

As it is seen in Fig.\ref{Bend-B-Fig1}, the parameters of
the disk stay unchanged for the first 2.5 turns ($t
\approx 10$ in our units). This time the modes are formed at the
linear stage of the bending instability evolution. After
$t \gee 10$ it changes to the nonlinear development of
the bending instability and the barycenter $\zeta(r)$
oscillates with larger amplitude in $z$-direction (see
Fig.\ref{Bend-B-Fig1}a). In the central regions of the
disk the amplitude $\zeta(r)$ rises up faster to its
maximum value and falls then down to the initial value
(see curves 1-5 in Fig.\ref{Bend-B-Fig1}a), whereas the
growth of $\zeta(r)$ occurs much slower at the outer
parts of the disk (curves 11-19). The rapid growth of the
velocity dispersion $c_z$ and of the vertical scale height $z_0$
begins after a time delay from $\zeta(r)$, see
Fig.\ref{Bend-B-Fig1}b and \ref{Bend-B-Fig1}c. The flare
emerges in the disk's central region and then propagates
toward its periphery. The value of $c_z/c_r$ rises up
mostly because of $c_z$ growth, and due to $c_r$
decrease as well (Fig.\ref{Bend-B-Fig1} d, e). The
azimuthal component of velocity dispersion $c_\varphi$
follows $c_r$ and the relation $c_\varphi\simeq c_r
\varkappa/2\Omega$ is valid almost everywhere. The
equation can be failed only either in a very thick disk or
in regions where $z_0/L\gee 0.4$. Note that
$c_r/c_\varphi < 2\Omega/\varkappa$ that means the
anisotropy is lesser in those regions because of the system
spherization. The decrease of $c_r$ and $c_\varphi$ is
a result of conversion of the kinetic energy of random
motion in the disk plane to the kinetic energy of
vertical motions. Note that the development of this bending
mode takes place in the disk which is axisymmetric in all
parameters.

\begin{figure}[!t]
{\includegraphics[width=0.49\hsize]{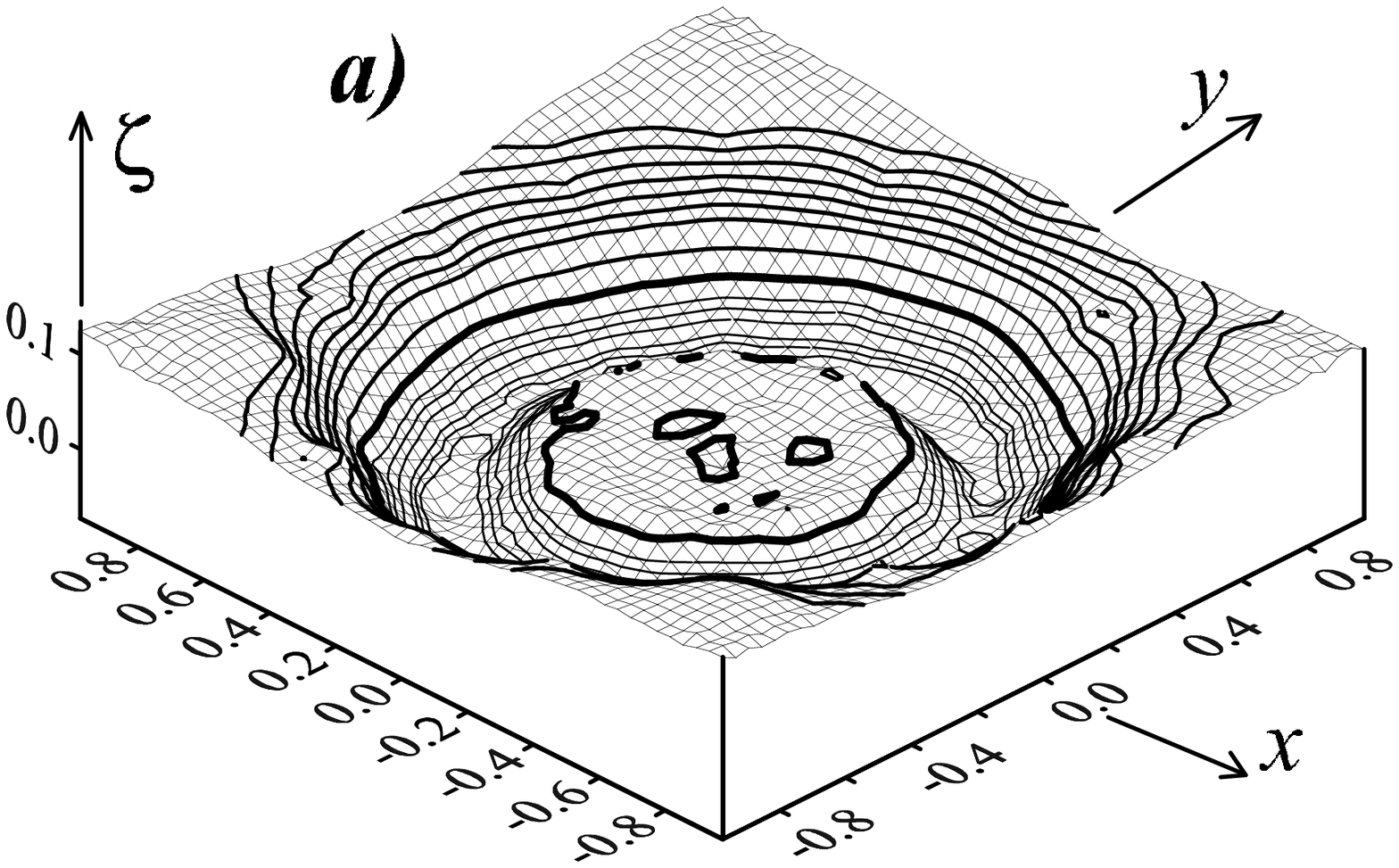}}
 \vskip -0.2\hsize \hskip 0.5\hsize
{\includegraphics[width=0.5\hsize]{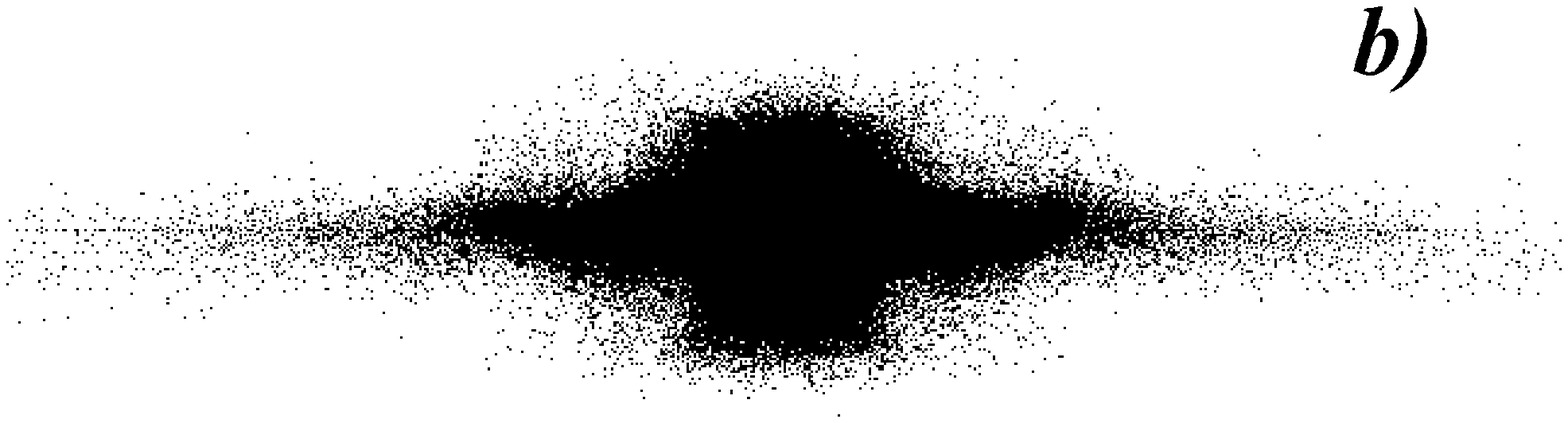}}\vskip
0.03\hsize
  \caption {
The shape of the stellar disk under the development of
the bending instability at the moment $t=25.4$ for the
model shown in Fig.1~a--f. Distribution of the
vertical coordinate of the barycenter $\zeta(x,y)$ are
shown in the $x-y$ plane
 (a). The bold solid line represents $\zeta=0$.
 b)~The edge-on view to the disk on the stage of the global
axisymmetric mode development. The vertical/plane aspect
ratio is increased 4 times.
}
\label{Bend-Fig-2}
\end{figure}

The value of $\zeta$ has different signs in disk's center
and periphery (see Fig.\ref{Bend-B-Fig1}a), what causes a
``mexican hat-like'' structure formation as a result of
this instability (Fig.2). The isolines of $\zeta$ have a
concentric shape. Up to the moment $t=25.4=6\tau$ the
internal ring and the periphery are shifted off the disk
plane toward opposite directions, whereas the central
region ($r\lee 0.4$) has returned to its initial
state.
 The disk vertical structure driven by the bending
mode $m=0$ is shown in Fig.2b. One can see a box-like
shape of the disk which is typical for all models where
the axisymmetric bending mode dominates and this feature
is not relevant to a bar.

A gradual change of the velocity dispersion $c_z$ and
$c_r$, and rotational velocity $V$ begins when the
amplitude of the vertical oscillations $\zeta$ grows
significantly. The vertical disk heating is accompanied by
increase of $c_z/c_r$ but starting with certain values
of $c_z/c_r$ the favourable conditions for developing of
the bending instability disappear. As a result, new
stable and thick disk forms. The characteristic time for this
process depends significantly on the model parameters and
is of order ten turns of the disks' outer regions. If the
initial distribution of $\alpha_z = c_z/c_r$ is
subcritical, the linear stage may take longer time due to
low increment of instability, and heating at
the nonlinear stage appears feeble.
Hence, the
amplitude of the bending mode and even the ability of its
emerging mostly depend on the initial radial
distribution of $c_z/c_r$.

The important consequence is that the final distribution
of $\alpha_z (r)$, as a result of bending perturbations,
depends on its initial value $\alpha_z (r, t=0)$. Thin
disks which have a low value of the scale height after the
action of the global bending instability get the larger
ratio $c_z/c_r$ than it would required to keep the disk
stable. The explanation lays in the essentially nonlinear
nature of the disk heating. The parameter
$c_z/c_r$ passes through the ``zone of stability'' and
stays off it and $c_z/c_r$ jumps over its marginal
threshold.

\begin{figure}[!t]
{\includegraphics[width=0.31\hsize]{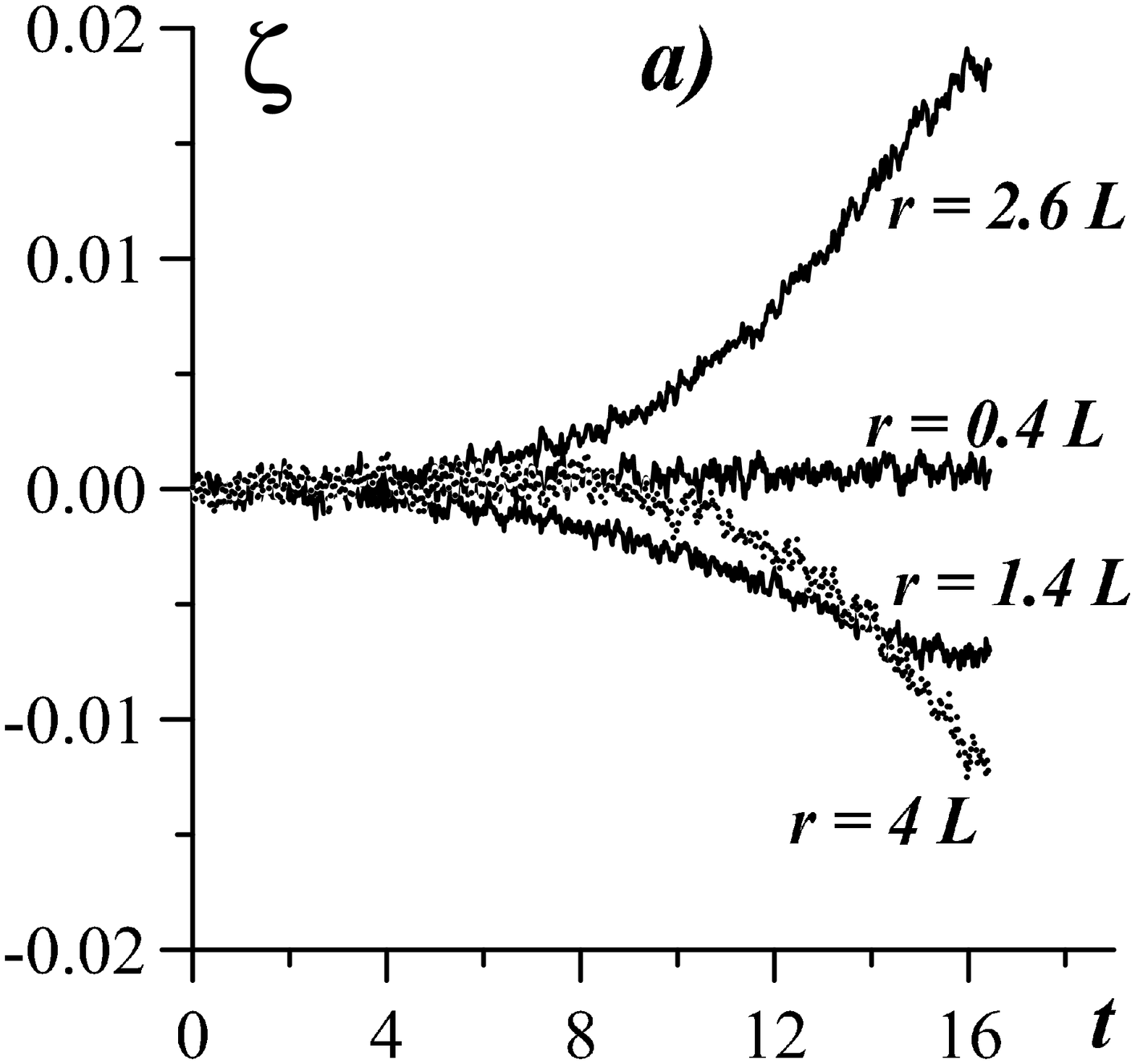}}
 \vskip -0.3\hsize \hskip 0.32\hsize
{\includegraphics[width=0.36\hsize]{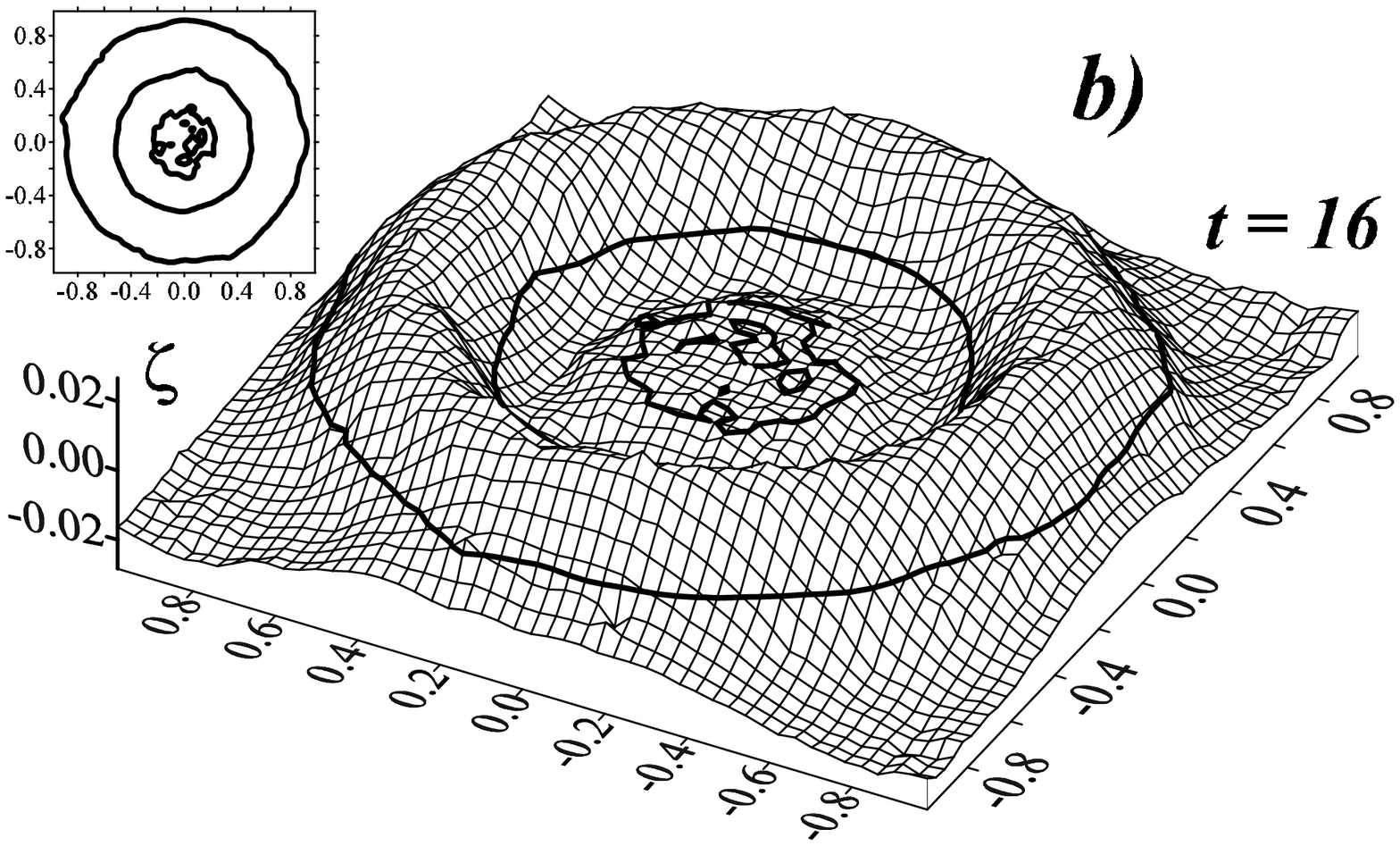}}
 \vskip -0.215\hsize \hskip 0.68\hsize
{\includegraphics[width=0.32\hsize]{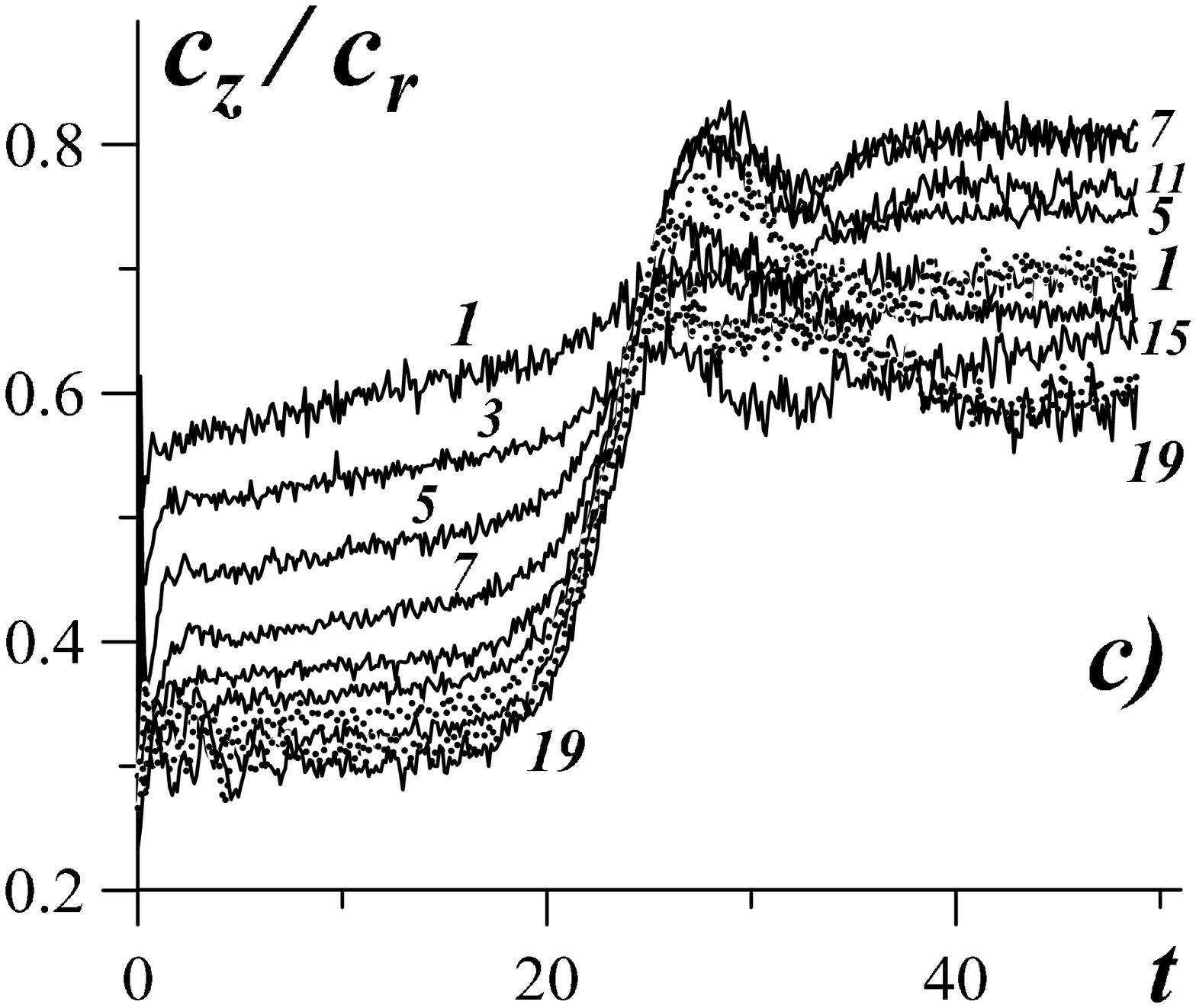}}
 \vskip 0.03\hsize
  \caption {
The model with $\mu=4$.
a) Initial moments of the vertical displacement evolution
at different distances from the center of the disk.
b) The shape of the stellar disk (i.e. distribution of
$z$-coordinate of its local barycenter through the disk).
The thick line corresponds to $\zeta=0$.
c) The flare of the disk without a halo but in presence of a bulge
($M_b=0.25\,M_d$, $b=0.2\,L$) after the development of the mode $m=0$.
In the regions of bulge the disk thickening is not significant.
The notation is kept the same as in Fig.1.
}
\label{Bend-Fig-3}
\end{figure}

If the disk is hot enough to provide the gravitational
stability against the bar mode and the mass of halo is
low ($\mu\lee 1$), the main cause of heating the
previously thin cold disk is the axisymmetric bending
mode (m=0) whereas the modes m=1,2 do not emerge at
all. Note that the mode m=0 might start forming far
from the center ($r\gee L$) not penetrating into the
central part of disk. It can happen by two ways: 1)
the initial disk has $c_z/c_r \ge \alpha_z^{crit}$ and
is stable in the central regions whereas the peryphery
of the disk is thin and unstable
(Fig.\ref{Bend-Fig-3}~a,~b); 2) when the galaxy has a
concentrated and massive bulge
(Fig.\ref{Bend-Fig-3}~c). In such models the bending
modes in the center have lesser amplitude and, in
general, a bulge, as well as a halo, plays a
stabilizing role. Hence, having all other equal
conditions, the disks of galaxies owing bulges are
thinner. Note that we don't consider dynamical models
for the bulge and as a correct approach a
non-stationary model for the bulge has to be
evaluated, hence this our result require further
investigation. Meanwhile we constrain our study in
\S~4 by the case of galaxies without large bulges.

\begin{figure}[!t]
\centerline{\includegraphics[width=0.7\hsize]{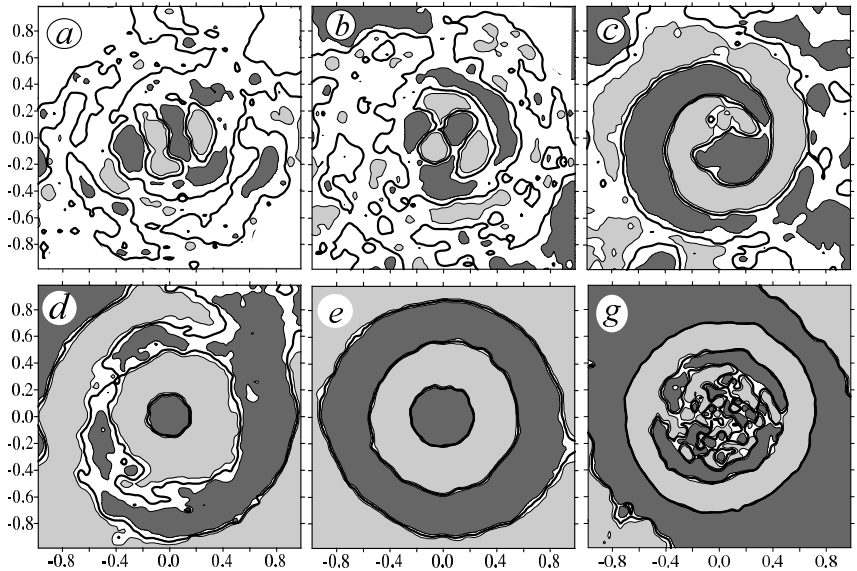}}
 \vskip -0.01\hsize
  \caption {
The structure of the vertical displacement $\zeta(x,y)$ in
the disk plane at different moments: a) $t=2.7$ -- the
mode $m=2$ developes itself in the central region, b)
$t=3.1$ -- $m=2 \rightarrow m=1$ reorganization, c)
$t=4.2$ -- a well-developed one-arm mode $m=1$, d)
$t=10.1$ -- $m=1 \rightarrow m=0$ reorganization, e)
$t=16.4$ -- a well-developed symmetric mode $m=0$, g) $t=51$
-- all perturbations in the center $r\lee L=0.25$ have
almost gone out. In the middle regions one can see the
axisymmetric perturbations and the mode $m=2$ takes place
at the outer parts of the disk.
}
\label{Bend-Fig-4}
\end{figure}

{\it The bending modes $m=1,2$. }
In the models with a massive halo ($\mu\gee 2$) the
axisymmetric bending modes $m=1$ and $m=2$ may develop
themselves and heat up the disk in the vertical direction.
The surface density of the disk remains axisymmetric. The evolution of
initially thin disk for $\mu=4$ is shown in Fig.1~g-i. In the
latter case the bending mode $m=2$ of saddle-type is being
developed firstly in central disk regions ($r\lee
2L$).
Fig.\ref{Bend-Fig-4} shows the distribution of $\zeta$ in the disk plane
for different moments of $t=0\div 40$.
The vertical heating and disk's flare
due to the mode $m=2$ is very modest in this case.
However, after $t\gee 3$ the non-linear stage of
one-arm asymmetric mode $m=1$ develops itself and the
vertical heating gets more significant. The third stage of
the heating begins at $t\gee 10$ when the mode $m=0$
begins to dominate in the disk. The growth rate of the
vertical dispersion $c_z$ is especially high at this time;
the disk scale height increases by a factor of 2 -- 3. The
example considered shows the process of transformation
between the bending modes and transition from the
asymmetric mode $m=2$ to the axisymmetric one.

As well as for the case of the spherical subsystem of low
mass, the favourable conditions for the emerging of the
global bending instability get worse and, starting with
some $\alpha_z(r)$, the disk warp does not develop itself
for at least 20 rotation turns if the initial $\alpha_z =
c_z/c_r$ (and $z_0(r)$ respectively) is high enough.

A stellar disk developing the bending instability with
$m=1,2$ is not in a quasy steady-state in the vertical
direction. The distribution of $c_z$ remains
axisymmetric, in a contrary to $z_0$. Hence, the
condition $c_z^2/z_0=const$ is failed and this
instability evolves rapidly. For the axisymmetric mode
the condition $c_z^2\propto z_0$ is fulfilled much
better except the very central region of the disk in
stages when the thickness rises significantly.

Apparently, an essential part of real stellar disks did
not undergo the heating due to the axisymmetric bending mode
$m=0$ because this mode would produce too thick disks. As an example,
in the experiment shown in Fig.1 ($\mu=1$) the disk scales
ratio $<z_0>/L\simeq 0.4$ (here $<z_0>$ denotes the average
thickness over the disk). The relation $c_z/c_r $ exceeds its critical
value for the final state of disk. The relative
thickness is less ($<z_0>/L\simeq 0.16$) for the case of
massive halo ($\mu=4$) after the relaxation of the
axisymmetric mode. On the other hand, there are galaxies
which have the ratio $<z_0>/L\le 0.15$
According to our assumptions either the very massive halos are required
for this case, or the axisymmetric bending mode $m=0$ was never developed
in those galactic disks.

\begin{figure}[!t]
\centerline{\includegraphics[width=0.73\hsize]{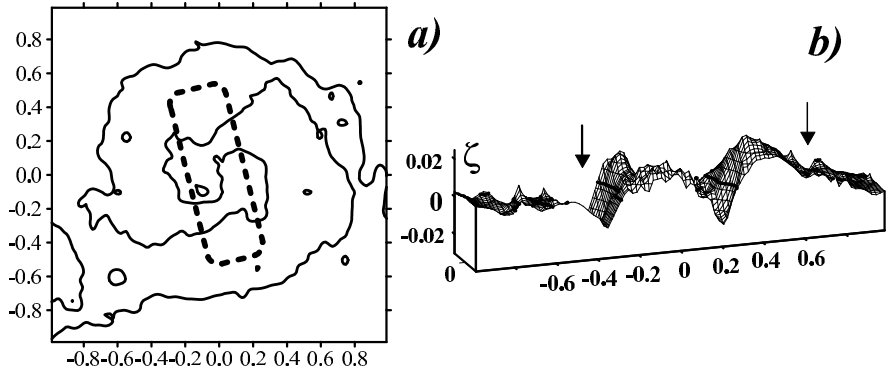}}
 \vskip -0.01\hsize
  \caption {
a) The distribution of $\zeta$ in the disk plane (dash-line
shows the bar position, the solid line marks $\zeta=0$).
b) The vertical profile of $\zeta$ taken along the bar.
The arrows mark the bar edges).
 }\label{Bend-Fig-5}
\end{figure}

 {\it Bendings of bar}.
Let's consider the experiments with rather light halo ($
\mu\lee 1.5 $). If the initial state of disk is gravitationally unstable,
a bar can be developed.
The bar formation is accompanied by its warps
\cite{Raha}. The bendings of a bar can emerge as a result of global
instability of the bar-mode during its initial stage when the bar forms
in initially thin cold disk. Fig.\ref{Bend-Fig-5} shows the bar bending
which increases the vertical velocity dispersion with time. The amplitude
of the bar warps falls essentially with increasing the bar thickness.
Let's stress that once the bar has been formed, it stops the further possible
developing of the global bending modes, thus it destroys the axisymmetric
mode $m=0$ first of all.

The bar formation occurs faster in the case of cold initial disk (for
small values of the Toomre stability parameter $Q_T$) and as a result,
the bar bending amplitude rises (Fig.\ref{Bend-Fig-5}).
% Let's note, that the conclusion about  warps of bar at its
% formation is not absolute.
If the initial disk was in a marginally subcritical state
(i.e. $c_r$ was just below the critical value which provides stability
against the global bar-mode), the bar forms very slowly and it is stable
against the bending perturbations.

\begin{figure}[!t]
{\includegraphics[width=0.47\hsize]{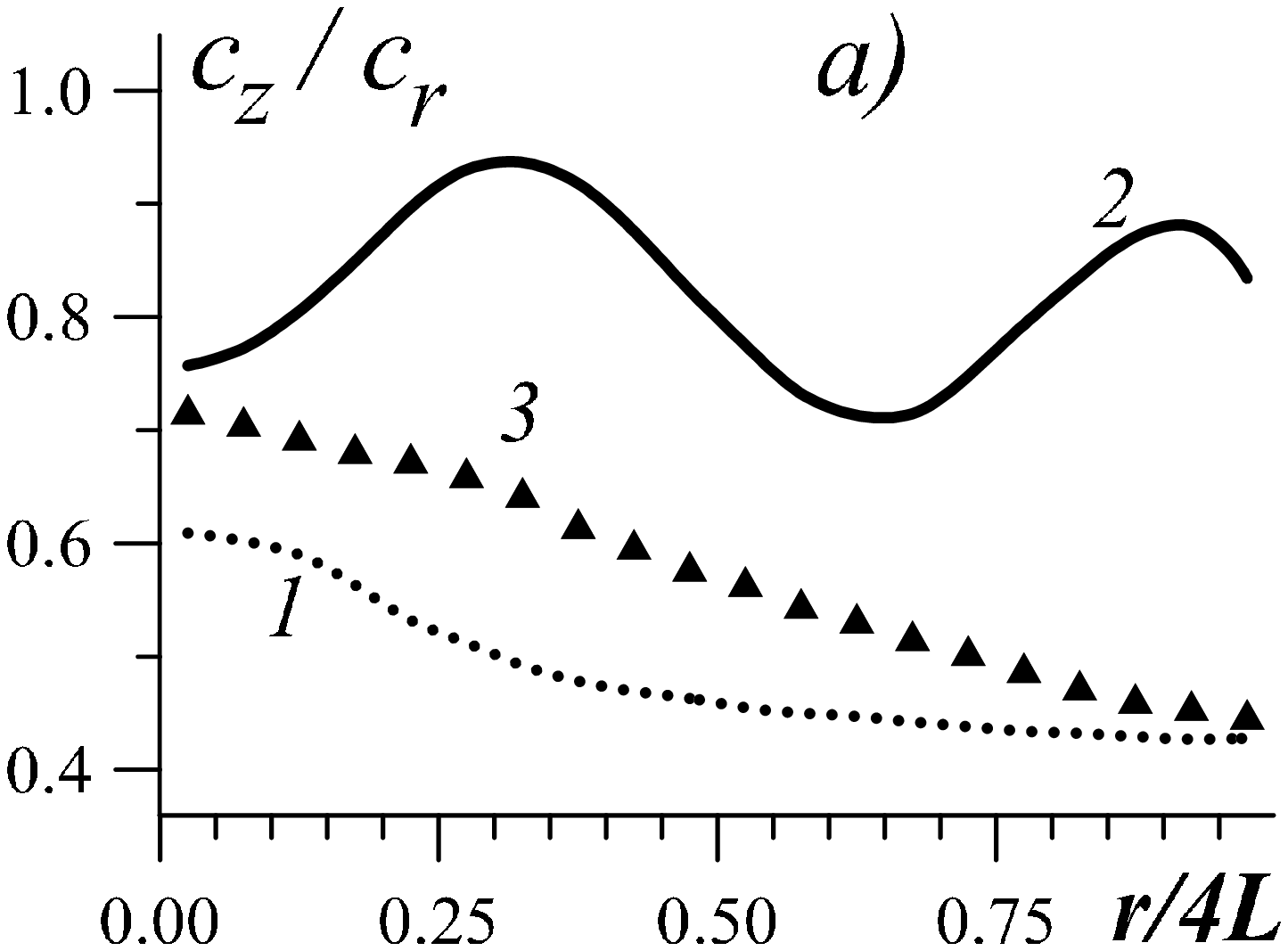}}
 \vskip -0.35\hsize \hskip 0.5\hsize
{\includegraphics[width=0.47\hsize]{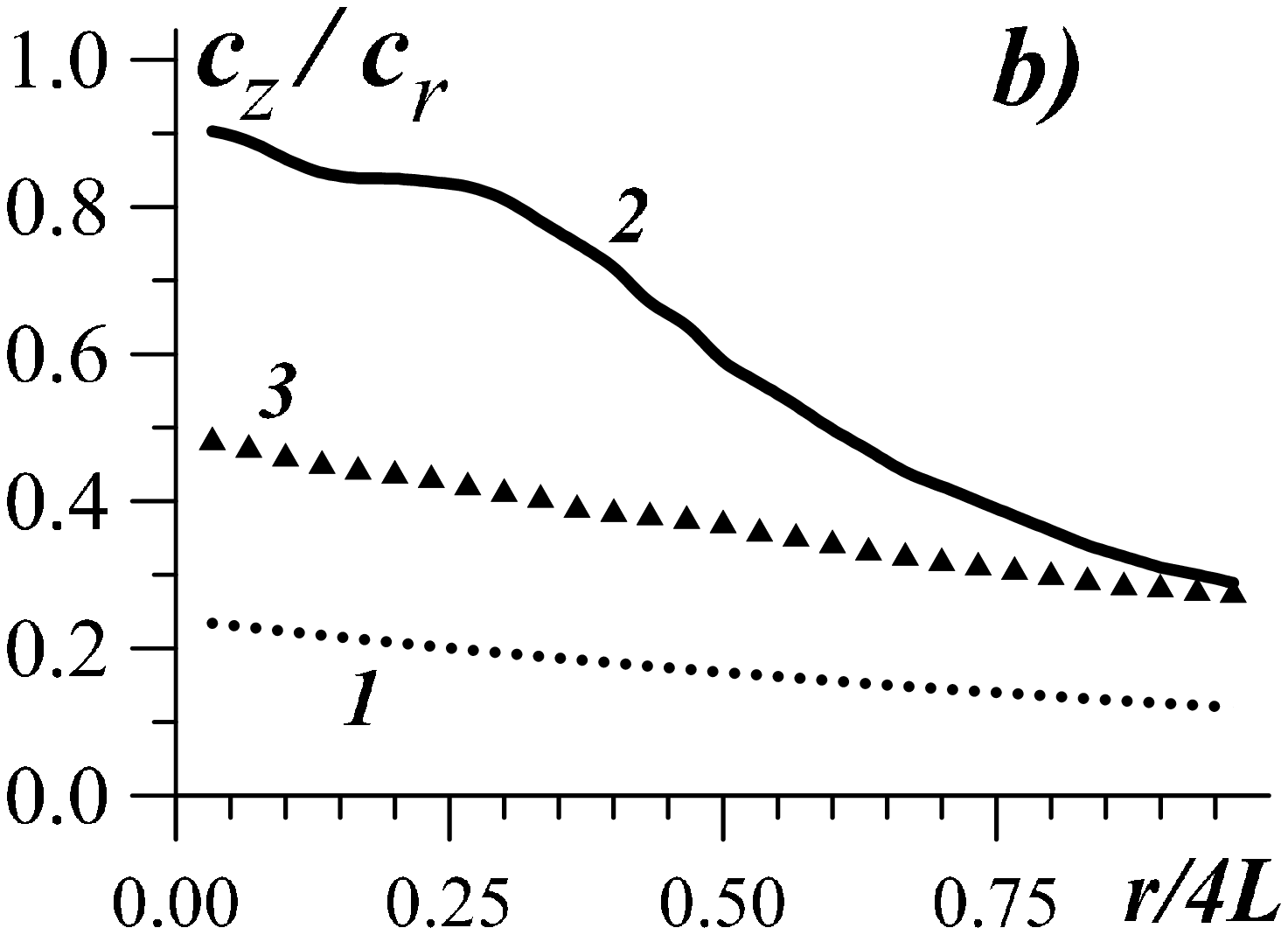}}
 \vskip -0.02\hsize
  \caption {
The radial distribution of $\alpha_z=c_z/c_r$ for: a)
$\mu=1$ (see Fig.1a--f), b) $\mu=4$ (see Fig.1g--i). 1 --
the initial distribution, 2 -- the final distribution, and
3 -- the critical level of $\alpha_z$.
 }\label{Bend-Fig-6}
\end{figure}

{\it The ratio $c_z/c_r$}.
With the similar values of parameters in our model, the
key parameter responsible for the stellar disks stability
is the ratio $\alpha_z=c_z/c_r$. To stabilize the global
bending perturbations in the case of low relative mass of
the spherical subsystem, the values of
$c_z/c_r$ have to be larger than $0.3-0.37$ what was
figured out from linear analysis in simple models
\cite{Poliachenko77, Araki-1986, Merritt}.
Let's consider bulgeless model with a moderate halo
$\mu=1$ (see Fig.\ref{Bend-Fig-6}). The initial (curve 1)
and final (curve 2) distributions of $\alpha_z(r)$ are
shown in Fig.\ref{Bend-Fig-6}a. The vertical heating of disk
originated from the axisymmetric bending mode development
is as strong that the averaged throughout the disk ratio
$<c_z/c_r>=0.82$. If the initial ratio $\alpha_z(r)$
follows the curve 3, then the global bending modes stay
stable.

Once the relative mass of halo rises gradually, the ratio $c_z/c_r$ needed
to marginally stabilize the bending perturbations (we will denote it as the
critical $\alpha_z$ ratio) is getting less. The initial (curve 1) and final
(curve 2) distributions of $\alpha_z(r)$ for the case of very massive halo
of $\mu=4$ are shown in Fig.\ref{Bend-Fig-6}b. The curve 3 shows the initial
distribution of $\alpha_z(r)$ needed to provide the stability against the
global bending modes. At the same time in the disk periphery
$0.27<c_z/c_r<0.37$.

A distinctive feature of the considered models at the
threshold of bending stability
is the non-uniformity of $c_z/c_r$ along the radius (see
Fig.\ref{Bend-Fig-6}). For the case of moderate halo
($\mu\lee 1$) and when the initial distribution of
$c_r(r)$ suppresses the bar instability, the critical
value of $\alpha_z^{crit}$ is a declining monotonous
function of $r$: its typical values range from $0.7\div 0.8$
at the central regions to $0.4\div 0.5$ at the periphery.

In both cases of large and small $\mu$, the ratio of
velocity dispersions $c_z/c_r$ falls exponentially
with radius and can be approximated as
$\alpha_z^{crit} \propto \exp(-r/L_\alpha)$ with scale
$L_{\alpha} \simeq (4-6) \cdot L$.

There are two reasons of decreasing the
disk thickness with the rising of the halo mass. At first,
the ratio $c_z/c_r$ needed to stabilize the bending
instability is lower.
On the other hand, the halo suppresses the
gravitational instability in the plane of disk, therefore
$c_r/V$ falls \cite{AVH2001, AVH2003}.
Note that in real galaxies several more factors such as
density waves, scattering on giant
molecular clouds, and tidal interactions heat up the disk
and increase its scale height. Then $\alpha_z^{crit}$ and
the corresponding disk thickness gives us the lower estimate
for the halo mass.

%*********************************************************
\section{ The results of dynamical modeling of edge-on galaxies}
%*********************************************************

In order to compare our model predictions with observations, we consider
seven spiral edge-on galaxies. Four of them, NGC~4738, UGC~6080, UGC~9442,
and UGC~9556 have no bulges and their structural parameters were derived
by \cite{dmbiz2002}. The
radial distribution of their stellar disks' thickness is available.
The galaxy UGC 7321 is interesting for us because of its superthin disk
and low surface brightness nature \cite{Matthews2000}. For two large and
nearby galaxies, NGC~891 and NGC~5170, the data on the radial component of
stellar velocity dispersion are available from published data
\cite{Bottema1987, Kruit1981, Bahcall1985, Shaw, Morrison1997,
Xilouris-1999, Sancisi1979, Broelis1991, Bottema1991}.
It enables us to incorporate those data together with the structural
parameters of disks.
The name of galaxy, the adopted distance $D$, the radial
scale length $L$, the averaged scale heights $<z_0>$ and
$<h_{z}>$ (corresponding to $sech^2$ and $exp$
distributions in the vertical direction respectively), and
the stellar disk maximum radius $R_{max}$ are shown in Table
\ref{Tabl1}.

\begin{table}[!t]
\begin{center}
\caption{Parameters of the edge-on galaxies}
\label{Tabl1}
\smallskip
\begin{tabular}{cccccc|c}
\hline
\hline
Name     &  $D$ & $L$      & $<z_0>$ & $<h_z>$ & $R_{max}$ & $\mu$\\
         &  Mpc & kpc      &   kpc   &   kpc   &  kpc &      \\
\hline
UGC 6080 & 32.3 & 2.9      & 0.69    & 0.48    & 9.9  & 0.57 \\
NGC 4738 & 63.6 & 4.7      & 1.3     & 0.7     & 19.2 & 0.7  \\
UGC 9556 & 30.6 & 1.5 (3.6)& 0.51    &  ---    & 9    & 1.1  \\
UGC 9422 & 45.6 & 3.5      & 0.80    & 0.51    & 14.6 & 0.8  \\
NGC 5170 & 20   & 6.8      & 0.82    &  ---    & 26.2 & 1.38 \\
UGC 7321 & 10   & 2.1      & 0.17$^a$& 0.14$^b$& 8.2  & 2.3  \\
NGC 891  & 9.5  & 4.9      & 0.98    & 0.49    & 21   & 0.89 \\
\hline
\end{tabular}
\smallskip
\end{center}
\vbox{\footnotesize
\noindent $^a$ the scale is shown for the disk's periphery
in the case of $sech(z/z_{ch})$.

\noindent $^b$ the scale is shown for the disk's central region.

\noindent Here $D$ is the distance to the galaxy
($H_0=75$~km~s$^{-1}$~Mpc$^{-1}$), $L$ is the exponential disk scale
length, $<z_0>$ is the averaged over the disk value of the vertical scale
height for $sech^2$ vertical profile, $<h_z>$ is the scale height averaged
over the disk for the exponential vertical profile, $R_{max}$ is the
maximum radius of the stellar disk.
}
\end{table}

% \hline \hline Name     & $D$   & $L$     & $\langle
% z_0\rangle$
% &$\langle h_{z}\rangle$ & $R_{max}$ & $\sigma_0^{disk}$ &$M_d$ &$\mu$\\
%   & Mpc  & kpc   & kpc     & kpc     & kpc  &
% $\frac{M_{\odot}}{pc^2}$&$10^{1
% \hline
% UGC 6080 & 32.3  & 2.9     & 0.69    & 0.48    &  9.9&1230&5.3 &0.57 \\
% NGC 4738 & 63.6  & 4.7     & 1.3     & 0.7     & 19.2&840&10.6 &0.7  \\
% UGC 9556 & 30.6  & 1.5 (3.6)& 0.51    &  ---    &  9 &720&1.4 (140) &1.1
% \\
% UGC 9422 & 45.6  & 3.5     & 0.80    & 0.51    & 14.6&1060&7.5 &0.8  \\
% NGC 5170 &  20   & 6.8     & 0.82    &  ---    & 26.2&490&12.8 &1.38  \\
% UGC 7321 & 10   & 2.1      & 0.17$^a$& 0.14$^b$& 8.15&250&0.62 &2.3   \\
% NGC 891  & 9.5   & 4.9     & 0.98    & 0.49     & 21 &650&9.1 &0.89   \\

\begin{figure}[!t]
{\includegraphics[width=0.47\hsize]{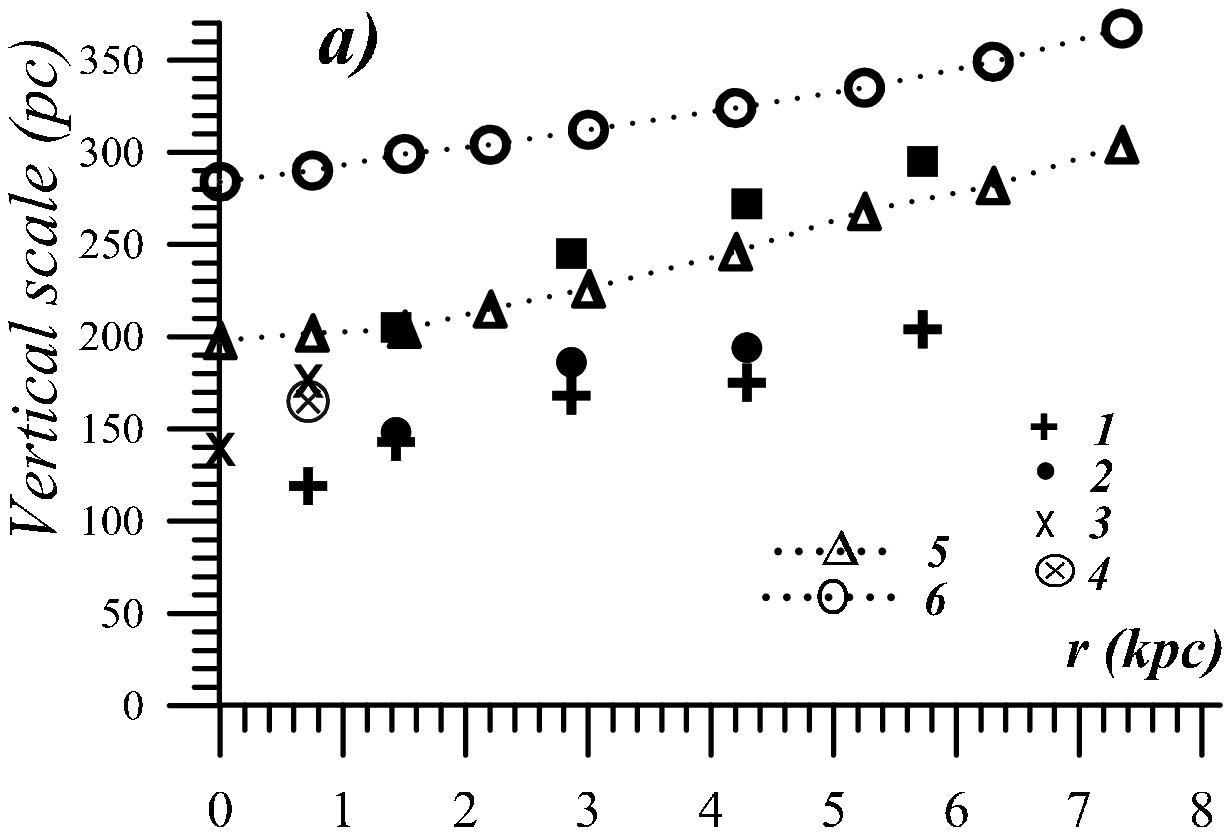}}
 \vskip -0.325\hsize\hskip 0.5\hsize
{\includegraphics[width=0.47\hsize]{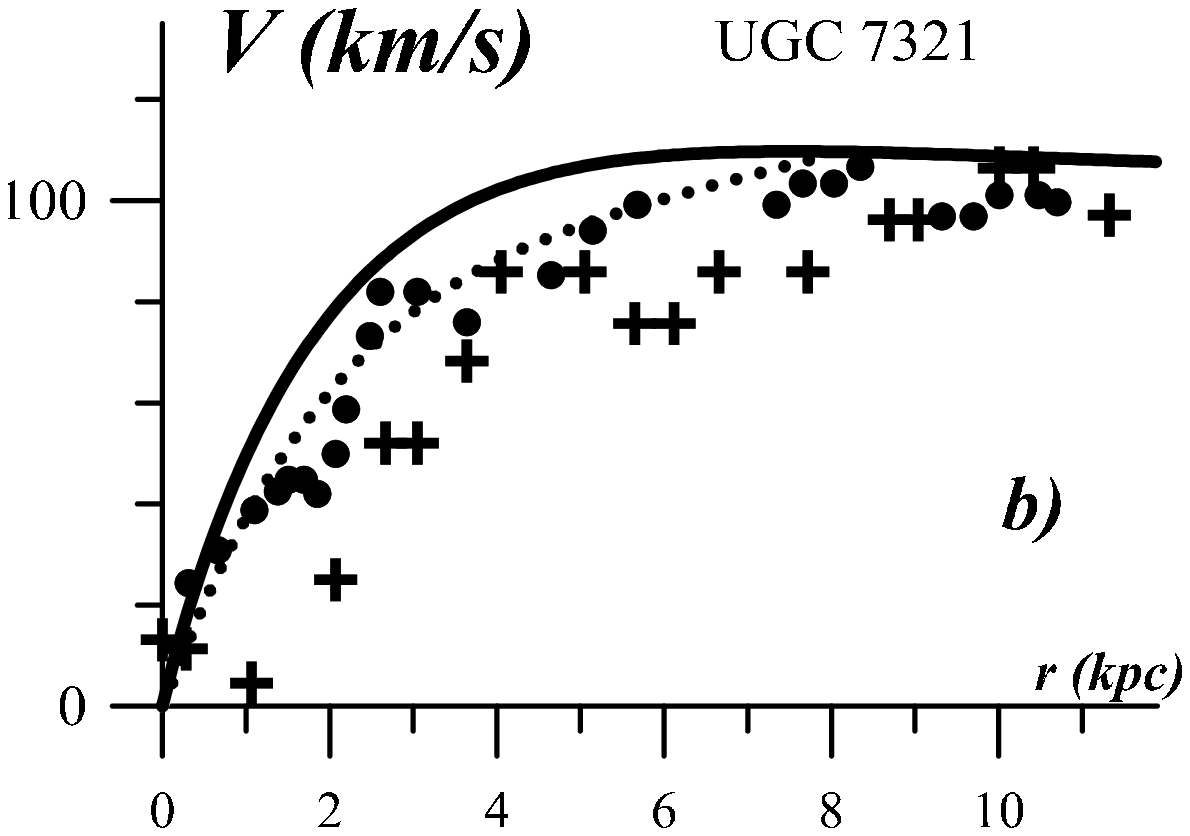}}
 \vskip -0.02\hsize
  \caption {
a) The radial distribution of the scale height in UGC~7321.
The observational data: 1 and 2 -- for the case
of ${\rm sech}(z/h_{ch})$ density distribution in the
vertical direction, {\it 3} and {\it 4} --- for the
exponential law $\exp(z/h_z)$.
b) The rotation curve of UGC~7321. The symbols $+$ and $\bullet$
represent the observational data for both sides from the center.
}
\label{Bend-Fig-7321}
\end{figure}

As an example, we consider the modeling of UGC~7321. The galaxy UGC~7321
reveals itself as a superthin \cite{Goal} and bulgeless
\cite{Matthews1999} disk galaxy. The disk thickness was derived by
\cite{Matthews2000}. We choose the systematic radial velocity of
394 km/s as that the maximum rotational velocity at the periphery
has the similar values ($V\simeq 100$~km/s) in both sides of the galaxy.
The curves 5 and 6 in Fig.14 are calculated for the models with
$\mu=2.3$ and $1.5$, respectively, inside of $r=R_{max}$. As
it is seen, the model with $\mu > 3$ gives a better
agreement with the observed radial distribution of the
vertical thickness. The best fit model with $\mu=2.3$ has
$M_d=0.62\cdot 10^{10}\,M_\odot$ and $M_h=1.4\cdot
10^{10}\,M_\odot$.

The evaluated best-fit parameters $\mu$ (spherical-to-disk mass ratios)
are shown in Table~1. Since five our galaxies have no bulges, and
the bulges of another two are relatively small, the values of $\mu$
represent the relative dark halo masses. As it can be seen from the Table~1,
the dark and luminous matter contributions are comparable inside the
stellar disk limits in our galaxies. The exception is the superthin
galaxy, UGC~7321, which has a massive dark halo and a low surface
brightness disk.

%zzzzzzzzzzzzzzzzzzzzzzzzzzzzzzzzzzzzzzzzzzzzzzzzzzzzzzzzz
\section{ Conclusions}
%zzzzzzzzzzzzzzzzzzzzzzzzzzzzzzzzzzzzzzzzzzzzzzzzzzzzzzzzz

\noindent 1. The development of the bending
instabilities in stellar galactic disks is studied
with the help of N-body numerical simulations. The
axisymmetric bending mode ($m=0$) is found to be the
strongest factor which may heat up stellar disks in
the vertical direction. If a bar formation was
suppressed, the role of ($m=1$) and ($m=2$) bending
modes in the disk thickening would be low. The most
significant thickening of the stellar disk occurs at
the initial non-linear stage of the bendings
formation. Once the bendings are destroyed, the
vertical heating becomes less effective. The lifetime
for the bending mode $m=0$ increases with the growth
of the relative mass of the spherical subsystem $\mu$.
We show that the initially thin disks increase their
thickness much more rapidly than those started from a
marginally subcritical state.
 It is important to
notice that the final disk thickness and $c_z $ depend
on its initial state and the definition of stability
boundary requires a special approach.

\noindent 2. The critical values of the ratio $\alpha_z^{crit}=c_z/c_r$
are considered as a function of the spherical subsystem parameters.
At the threshold of stability, the value of $\alpha_z^{crit}(r)$ falls
with the distance to the center. The value of $\alpha_z^{crit}$ can be
twice less at the periphery in comparison with its central value.

%-We approximate dependence $c_z/c_r\propto \exp(-r/L_\alpha)$
%-with the scale $L_\alpha\simeq (5-6) \cdot L$.

\noindent 3. The averaged relative disk scale height $<z_0>/L$
falls when the relative halo mass increases. We use this kind of
relation to estimate the mass of the spherical subsystem for
edge-on galaxies.

\noindent 4. We conduct N-body modeling for seven edge-on galaxies based
on published rotation curves and surface photometry data (for two galaxies
we incorporate the data on the observed stellar radial
velocity dispersion in addition). The relative mass of the spherical
subsystem (dark halo in most cases) is inferred for all the galaxies.
The evaluated mass of the dark halo in our galaxies is of order of their
disk's mass. As an exception, the superthin LSB galaxy UGC~7321 own
the dark halo which contains more than 2/3 of overall galaxy mass.

\acknowledgements
This work is supported by the Russian
Foundation for Basic Research through the grants RFBR
04-02-16518, 04-02-96500 and by the Technology Program
``Research and Development in Priority Fields of Science
and Technology'' (contract 40.022.1.1.1101).

\end{document}